# Non-destructive electron microscopy imaging and analysis of biological samples with graphene coating

Jong Bo Park,[1,‡] Yong-Jin Kim,[1,2,‡] Je Min Yoo,[1] Youngsoo Kim,[1,6] Seong-Min Kim,[5] Sang Jin Kim,[1] Roman Gorbachev,[2] I. I. Barbolina,[3,4] Myung-Han Yoon,[5] Byung Hee Hong,[1,*] Konstantin S. Novoselov[2,*]

[1]*Department of Chemistry, College of Natural Science, Seoul National University, 599 Gwanak-ro, Gwanak-gu, Seoul 151-747, Korea*

[2]*Department of Physics & Astronomy, University of Manchester, Manchester M13 9PL, UK*

[3]*Faculty of Life Sciences, Michael Smith Building, Oxford Road, Manchester, M13 9PT, UK*

[4]*Technical Service Consultants Ltd, Microbiology House, Fir Street, Heywood OL10 1NW, Lancashire, UK*

[5]*Department of Materials Science and Engineering, Gwangju Institute of Science and Technology (GIST), Gwangju 500-712, Korea.*

[6]*Department of Physics and Astronomy, Seoul National University, 599 Gwanak-ro, Gwanak-gu, Seoul, 151-747, Korea*

‡These authors contributed equally to this work.

**In electron microscopy, charging of non-conductive biological samples by focused electron beams hinders their high-resolution imaging. Gold or platinum coatings have been commonly used to prevent such sample charging, but it disables further quantitative and qualitative chemical analyses by energy dispersive spectroscopy (EDS). Here we report that graphene-coating on biological samples enables non-destructive high-resolution imaging by scanning electron microscopy (SEM) as well as chemical analysis by EDS, utilizing graphene's transparency to electron beams, high conductivity, outstanding mechanical strength, and flexibility. We believe that the graphene-coated imaging and analysis would provide us a new opportunity to explore various biological phenomena unseen before due to the limitation in sample preparation and image resolution, which will broaden our understanding on the life mechanism of various living organisms.**

Comprehensive understanding of biological objects – their chemical, physiochemical and biological characteristics – can be effectively achieved through electron microscopy (EM) analysis[1-4], preferably without any fixation or auxiliary surface treatment. Scanning electron microscope (SEM) and transmission electron microscope (TEM) are increasingly more employed as they provide direct imaging of specimen's morphological structures[5] with high-resolution. In addition, the unique interaction between electron beams and specimen enables various physical and chemical analyses such as energy dispersive spectroscopy (EDS), electron probe micro analysis (EPMA), and electron energy loss spectroscopy (EELS)[6,7]. Nevertheless, charge accumulation on non-conductive surface by electron beams has always hindered EM-mediated biological studies as it distorts the morphological and chemical characteristics of the specimens[8-12]. Thus, various metal deposition methods have been employed to dissipate the charge in the non-conducting specimens[13-15]. However, the relatively thick metal coatings hamper from studying sample's fine surface structure at nanometer scale, because the surface features smaller than metal grain size (~10 nm) cannot be imaged properly. In addition, X-ray fluorescence signals required for EDS analysis are screened by metal layers[16]. Furthermore, it is usually difficult to use the metal-coated samples for further analyses such as TEM that requires electron-transparency. Here we report that graphene-coating on biological samples enables non-destructive high-resolution imaging by scanning electron microscopy (SEM) as well as chemical analysis by EDS, utilizing graphene's transparency to electron beams, high conductivity, outstanding mechanical strength, and flexibility[17-20]. We believe that the graphene-coated imaging and analysis would provide us a new opportunity to explore various biological phenomena unseen before due to the limitation in sample preparation and image resolution, which will broaden our understanding on the life mechanism of various living organisms.

A recent progresses in large scale synthesis of high quality graphene films using chemical vapor deposition (CVD) methods have widened its potential in practical device applications as well as unique interests in basic scientific researches[21-24]. The feasibility of the large scale fabrication of continuous graphene films as well as easy transfer onto diverse biological objects opens up a unique opportunity to create new hetero-interfaces or interfaces with non-conducing biological samples. As demonstrated in a recent work[25], the *in-situ* high-resolution EM imaging of nanocrystal growth has been achieved by using graphene liquid cells to encapsulate nanoscale materials as well as their environment (*i.e.* liquid) and separate them from the vacuum environment. In this regard, graphene mediated coating on biological samples can provide high-resolution EM imaging and chemical analysis due to the excellent electron and heat flow thorough the graphene and electron-transparency. Here, with taking all these advantages of graphene films, we have employed continuous graphene films as coating for biological samples and exploited them for non-destructive high-resolution electron microscopy imaging and chemical analysis.

As schematically displayed in Fig. 1, the unique feasibility and availability of continuous graphene films at large scale enables the conformal coating of biological objects including leaves, ants, spiders, neuron cells, and *Escherichia coli* (*E. coli*), whose sizes range from several centimeters down to few micrometers. Atomically-thin and electrically-conducting graphene membranes were prepared on non-conducting biological surface by isolating graphene films from copper (Cu) foils after CVD growth, followed by conformal coating onto biological samples as illustrated in Fig. 1C. Compared to other conventional sample preparation methods including fixation and metal sputter coating (Figs. 1a and 1b), the present method based on graphene coating is relatively simple, bio-friendly, and non-destructive, which is particularly advantageous for preserving samples for further experiments.

Monolayer graphene film was prepared on high-purity Cu foil using CVD method (please see supplementary materials). Continuous graphene films coated with protecting polymer layers (*i.e.* poly(methyl methacrylate) (PMMA)) can be isolated from Cu foils and transferred to a desired surface after wet chemical etching[23]. The number of graphene layers was controlled by repeating this transfer process. We found that triple-layered (3-layer) graphene films provide both electron transparency and mechanical stability optimized for SEM analysis even without PMMA supports (Fig. S1). For actual SEM imaging, the biological specimens were cleaned and transferred onto a metallic sample stage to facilitate electron discharge. The 3-layer graphene sheets were then transferred on top of the biological specimen by scooping from bottom side, followed by drying in a desiccator.

To demonstrate the advantages of using graphene membrane for SEM imaging, we have selected several representative biological specimens (ants, bee's wings, water fleas, and *E. coli*) that are different in terms of size, surface hardness, and morphology. The 3-layer graphene mostly covered these millimeter to sub-micron sized samples, and only shows minor fractures caused by mechanical deformation around needle-like structures (Fig. 2c). In contrast to CVD graphene coating, graphene oxide (GO) and reduced graphene oxide (rGO) coating methods resulted in inhomogeneous coating possibly due to their poor mechanical strength and difference in hydrophobicity (Fig. S2). The high-magnification FE-SEM images of a graphene-coated ant clearly show not only unique micro-patterns but also nano-pores as small as 40 nm (Fig. 2b) that are invisible in Pt-coated samples (Fig. 2i). Such fine and clear observation of the surface structures implies that the adhesion between graphene and the sample (mostly by *van der Waals* interaction) is strong enough to maintain its morphology[26] and stable up to acceleration voltage of 20 keV (Fig. S3). The needle-like structures on bee's wings result in punctures on graphene, but the surface still shows conformal graphene coating that enables stable SEM imaging (Fig. 2c).

We also performed SEM imaging on a 1.5 mm long water flea (*Daphnia pulex*) covered with graphene membranes. High-magnification SEM images of the water flea (area P1 in Fig. 2d) clearly display the unique features of a water flea on its dorsal carapace (Fig. 2e). Interestingly, the graphene film mostly covers the needle-like surface on its antenna without much tearing (Fig. 2f). The advantages of graphene coating compared with a conventional metal coating method was demonstrated under identical conditions (Figs. 2g-i). We observed that the bare gaster surface of an ant is strongly charged even at low acceleration voltages (< 2 keV) (Fig. 2G), and the bare eye surface is immediately burning at 5 keV, while the graphene-coated area doesn't show any damage even with high acceleration voltages up to 20 keV (Figs. 2g and 2h).

Unlike the above mentioned hard-surfaced insects, soft biological objects such as living cells and bacteria need to be treated with aldehyde fixation, osmium tetroxide staining, and critical point drying processes for SEM imaging, which often distorts the sample contents and disables further qualitative or quantitative chemical analyses. In this regard, the simple graphene-coating method is advantageous because biological samples close to their native structures can be imaged and preserved for further analyses. We demonstrate that a common bacteria, *E. coli*, cultured in a liquid medium can be imaged after graphene-coating that protects *E. coli* from sudden vacuum drying as well as e-beam damage (Fig. 2j-l) in SEM. Another graphene layer on bottom side was used to seal the liquid environment by $\pi-\pi$ interaction with top graphene layers. We expect that further combination with microfluidic methods would enable the real-time live imaging of bacteria and cells by electron microscopy in the future.

It is difficult to characterize the structures of deoxyribonucleic acids (DNAs) by SEM because of their vulnerability to e-beam radiation at vacuum. Figs. 2m-o demonstrate that DNA supercoiled structures can be visualized with graphene-coating that maintains native-like structures of DNAs by encapsulating surrounding water layers. We suppose that strong contrast in SEM images comes from charging of the surrounding water layers rather than DNA itself, which can be evidenced by no contrast in the cracked region without water layers (Fig. 2n).

Furthermore, we compared the performances of graphene-coating and Pt-coating methods in chemical analysis by energy dispersive spectroscopy (EDS) (Fig. 3). All the experimental conditions and parameters including spot sizes and signal collection time were identical. The results show that EDS signals from graphene-coated samples (Fig. 3a) are 2~3 times stronger than Pt-coated samples (Fig. 3b), which facilitates the qualitative and quantitative chemical analyses on, for example, nitrogen-containing chitin (from ants) and oxygen-rich cellulose (from leaves). The non-destructive analysis enabled by graphene coating is particularly effective for element-specific EDS mapping. The water flea sample was fed on sub-25 nm cerium oxide nanoparticles ($CeO_2$ NPs) to stain its digestive pathway (please see supplementary materials). The $CeO_2$ NPs are clearly visualized in the Ce-selective EDS mapping of the graphene-coated ant, while the Pt-coated ant doesn't shows clear EDS signals (Figs.3d and 3e). The other EDS analyses also indicate that the graphene-coated method is superior to Pt-coating in terms of signal intensity. We attribute the signal reduction in Pt-coated samples to the absorption and scattering of incident electrons and X-ray fluorescence radiation by thick Pt layers, which will be further discussed in Fig. 4.

The outstanding performance of atomically thin graphene membrane as protective coating for EM analysis was theoretically confirmed by Monte Carlo simulations (please see supplementary materials for detailed methods). The 1 nm graphene-coated chitin (Gr/Chitin) was compared with 10 nm Pt coated chitin (Pt/Chitin). As seen in the electron trajectory images (Fig. 4a), incident electrons can easily pass through the thin graphene and penetrate deep into the chitin layer, while the Pt layer blocks electron penetration because of the large nucleus radius (i.e. large electron scattering cross-section) of Pt (Z=78) (see supplementary Fig. S5 for cross-section calculations). As the acceleration voltage increases from 2 keV to 10 keV, the maximum penetration depth of electrons increased from 140 nm to 2,250 nm for Gr/Chitin. This indicates that graphene membrane is more transparent to lower accelerating voltages. On the contrary, the electrons irradiated to Pt/Chitin show less penetration depths with larger scattering angles. The amount of penetrated electrons is directly related to the X-ray signals (Fig. 4c), resulting in large difference in X-ray absorption intensity between Gr/Chitin and Pt/Chitin. Fig. 4b shows the cross-section profiles of energy loss along the simulated electron trajectories[27], which is related to the intensity of EDS signals. From 2 keV and 5 keV, most of the energy loss happens inside the 10 nm Pt layer. Major energy loss still occurs near the Pt layer at 10 keV. On the other hand, energy loss in Gr/Chitin mostly takes place inside the chitin layer even at 2 keV, indicating that the graphene is almost free from electron energy loss and background EDS signals. Fig. 4C shows the depth profiles of Rho-Z X-ray intensity of carbon for Gr/Chitin (red) and Pt/Chitin (grey). As the accelerating voltage increases from 2 keV to 10 keV, the total X-ray intensities for Pt/Chitin and Gr/Chitin increased from 6 to 259 and from 87 to 470, respectively (see also supplementary Fig. S8 for the simulations for nitrogen and oxygen analysis). This indicates that graphene is superior to Pt for protective coating for EDS analysis. The experimental EDS spectra with varying accelerating voltages coincide with the simulation results (Fig. S9).

In conclusion, we demonstrated that graphene's outstanding mechanical strength, conductivity, flexibility, and transparency to electron beams enable the simple and non-destructive imaging and analysis of various biological samples with high resolution that can be hardly achieved in bare or metal-coated samples. The graphene coating effectively prevent charge accumulation by spreading e-beam induced charges and heats over large surface area, and the high mechanical strength and flexibility of graphene allows conformal coating by excellent adhesion with various biological interfaces. We believe that the graphene-coated imaging and analysis would provide us a new opportunity to explore various biological phenomena unseen before due to the limitation in sample preparation and image resolution, which will broaden our understanding on the life mechanism of various living organisms.

**Supplementary Information** accompanies the paper.

**Acknowledgements** This research was supported by the Basic Science Research Program (2011-0006268, 2012M3A7B4049807), Converging Research Centre Program (2013K000162), the Global Frontier R&D Program on Centre for Advanced Soft Electronics (20110031629) and the Global Research Lab (GRL) Program (2011-0021972) through the National Research Foundation of Korea funded by the Ministry of Science, ICT & Future, Korea. This research was also supported by Inter-University Semiconductor Research Center (ISRC) at Seoul National University. KSN is grateful to the Royal Society, European Research Council and EC-FET European Graphene Flagship for support.

**FIGURE LEGENDS**

**Figure 1. Schematic illustration of various biological objects in different scales and coating methods for EM analysis. a** and **b**, Conventional coating methods of non-conducting biological samples. Soft biological samples such as cells and bacteria require complicated coating processes including aldehyde fixation, osmium tetroxide fixation, dehydration, critical point drying, staining, metal coating, etc. Hard-surfaced biological samples such as insects and plants are usually coated with Au, Pt by vacuum sputtering. The metal coating is simple, but it disables high-resolution imaging and analysis. **c**, Simple coating process using graphene floating on water surface. The ambient drying process allows the conformal coating of graphene on sample surface.

**Figure 2. SEM images of various biological samples covered with graphene films. a** and **b**, Low- and high-magnified SEM images of a graphene-coated ant. **c**, SEM images of grapheme-coated bee's wing, where about 30 μm sized needle-like structures are uniformly arrayed. **d-f**, Low- and high-magnified SEM images of a graphene-coated water flea, respectively. The graphene film exhibits conformal contact with the non-flat surfaces of biological samples. Acceleration voltages for A to F, 2 keV. **g-i**, Representative SEM images showing the comparison between bare, graphene-coated, and Pt-coated samples. The graphene coating enables the stable SEM imaging of sub-10 nm features on the surface, while the Pt-coated sample shows distorted morphology covered with Pt nanoparticles. **j**, Optical microscopic image of graphene-coated *E. coli.* **k** and **l**, Low- and high-magnification SEM images corresponding to j obtained with acceleration voltage at 2 keV. **m** and **n,** Low and high magnification SEM images of graphene-coated DNAs from *E. coli,* respectively. **o**, SEM image of DNAs without the graphene coating (bare) on Si wafer at 2 keV.

**Figure 3. SEM and EDS analyses of biological samples coated with graphene and Pt. a-c**, EDS spectra of a graphene-coated ant, a Pt-coated ant, and a graphene-coated leaf, respectively. Acceleration voltages, 10 keV. **d** and **e**, Representative SEM and EDS mapping images of a graphene-coated water flea and a Pt-coated water flea. Acceleration voltages, 20 keV. Scale bars, 200 µm. The corresponding EDS spectra are shown in supplementary fig. S6.

**Figure 4. Simulation results of the Pt/Chitin and the Gr/Chitin with 2, 5, and 10 keV accelerating voltages by Monte Carlo Calculation. a**, Electron trajectories of Pt/Chitin and Gr/Chitin. Blue lines are the trajectories of electron absorbed in the samples. Red lines are the trajectories of back scattered electron that would escape sample surface. **b**, Energy distribution contour mapping images of Pt/Chitin and Gr/Chitin. Dark red and dark blue are 100 and 0, respectively. **c**, Phi Rho-Z X-ray absorbed intensity of carbon (Phi = X-ray generation function, Rho-Z = a way of plot generation per unit density). Red and dark grey areas show the total intensities of Gr/Chitin and Pt/Chitin, respectively. The absorbed intensity of the Gr/Chitin are 87, 290, and 470 at 2 keV, 5 keV, and 10 keV, respectively. The intensity of Pt/Chitin are 6, 116, and 259 at 2 keV, 5 keV, and 10 keV, respectively.

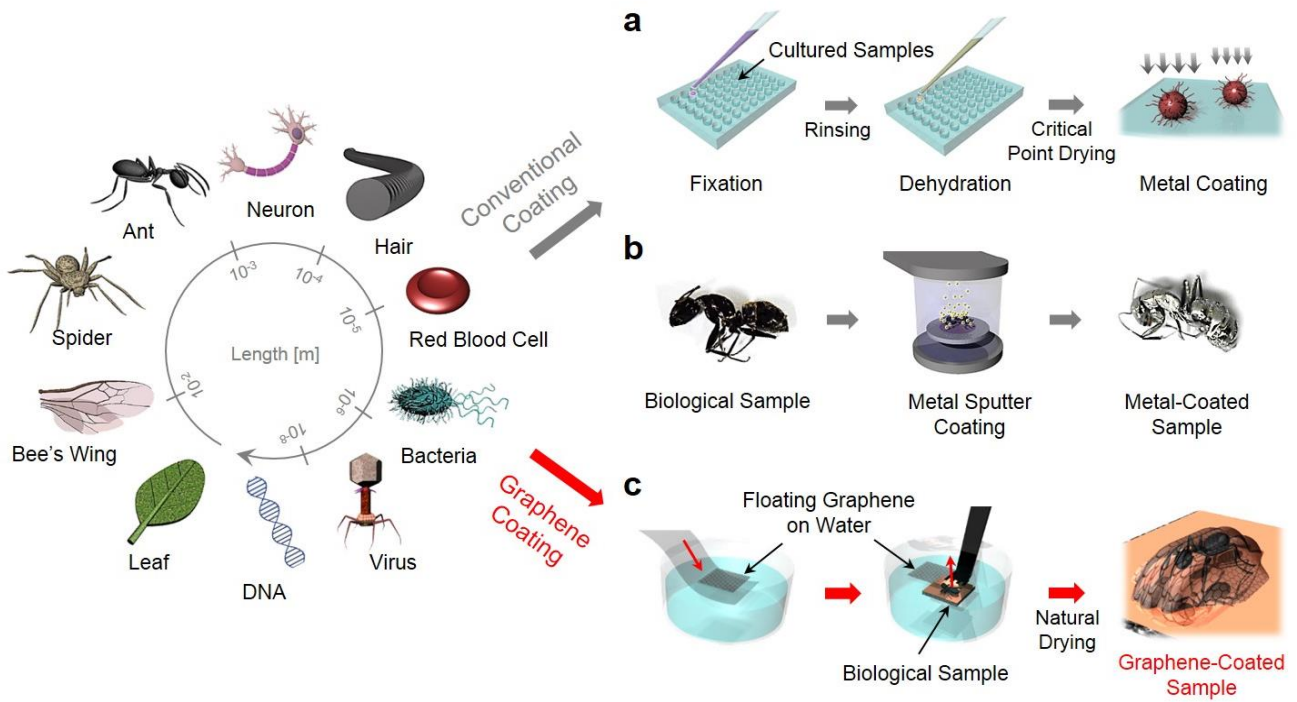

**Figure 1. Schematic illustration of various biological objects in different scales and coating methods for EM analysis. a** and **b**, Conventional coating methods of non-conducting biological samples. Soft biological samples such as cells and bacteria require complicated coating processes including aldehyde fixation, osmium tetroxide fixation, dehydration, critical point drying, staining, metal coating, etc. Hard-surfaced biological samples such as insects and plants are usually coated with Au, Pt by vacuum sputtering. The metal coating is simple, but it disables high-resolution imaging and analysis. **c**, Simple coating process using graphene floating on water surface. The ambient drying process allows the conformal coating of graphene on sample surface.

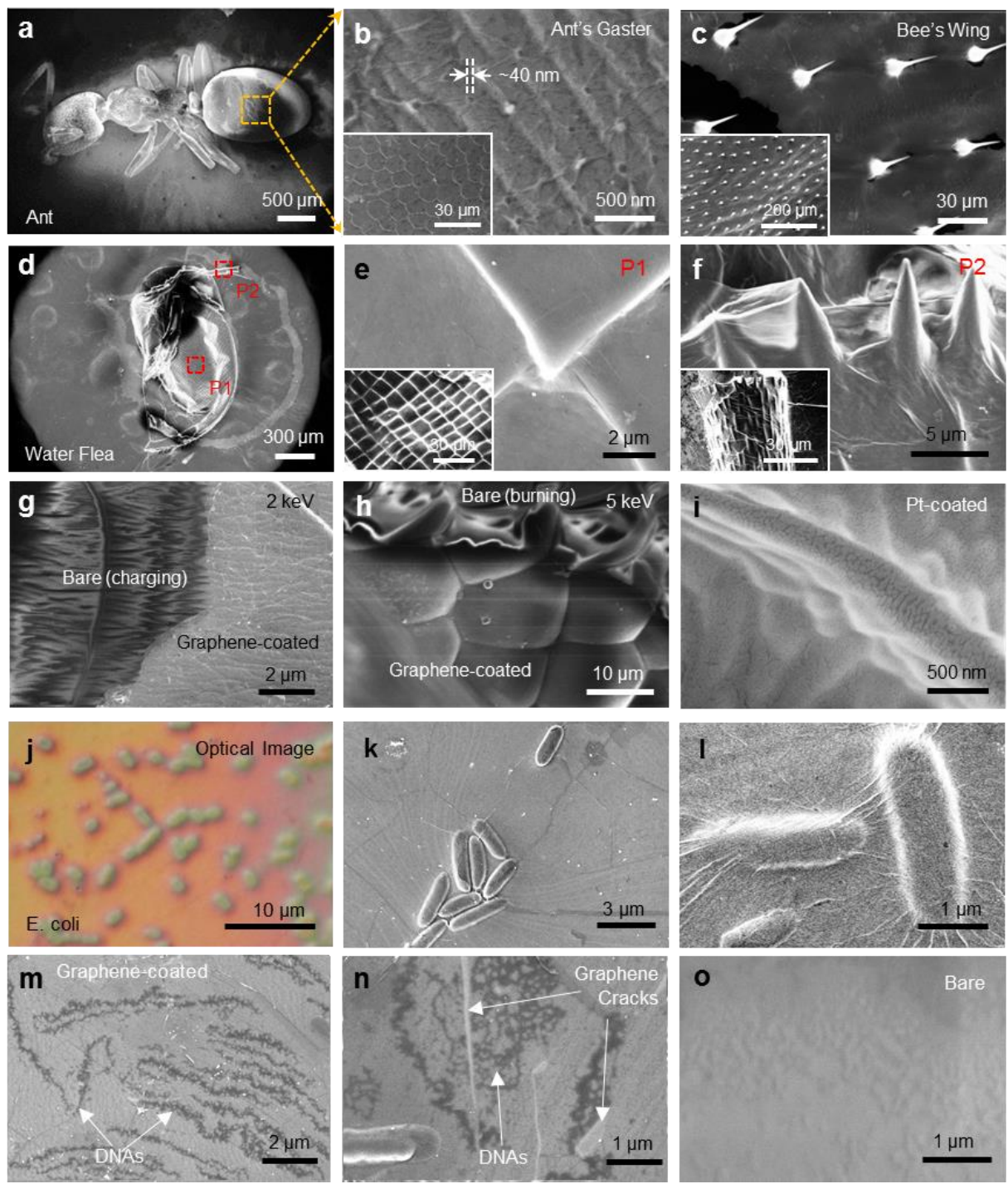

**Figure 2. SEM images of various biological samples covered with graphene films. a** and **b**, Low- and high-magnified SEM images of a graphene-coated ant. **c**, SEM images of grapheme-coated bee's wing, where about 30 μm sized needle-like structures are uniformly arrayed. **d-f**, Low- and high-magnified SEM images of a graphene-coated water flea, respectively. The graphene film exhibits conformal contact with the non-flat surfaces of biological samples. Acceleration voltages for A to F, 2 keV. **g-i,** Representative SEM images showing the comparison between bare, graphene-coated, and Pt-coated samples. The graphene coating enables the stable SEM imaging of sub-10 nm features on the surface, while the Pt-coated sample shows distorted morphology covered with Pt nanoparticles. **j**, Optical microscopic image of graphene-coated *E. coli*. **k** and **l**, Low- and high-magnification SEM images corresponding to j obtained with acceleration voltage at 2 keV. **m** and **n,** Low and high magnification SEM images of graphene-coated DNAs from *E. coli,* respectively. **o**, SEM image of DNAs without the graphene coating (bare) on Si wafer at 2 keV.

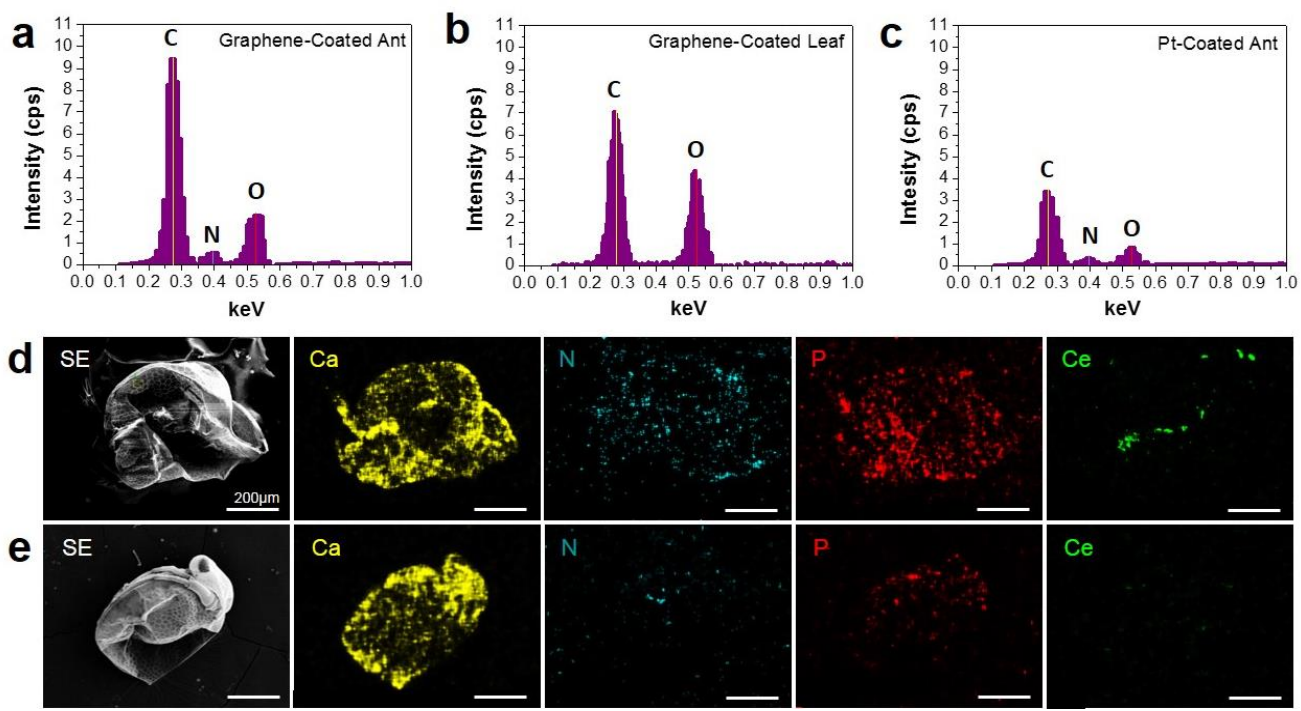

**Figure 3. SEM and EDS analyses of biological samples coated with graphene and Pt.**
**a-c**, EDS spectra of a graphene-coated ant, a Pt-coated ant, and a graphene-coated leaf, respectively. Acceleration voltages, 10 keV. **d** and **e**, Representative SEM and EDS mapping images of a graphene-coated water flea and a Pt-coated water flea. Acceleration voltages, 20 keV. Scale bars, 200 μm. The corresponding EDS spectra are shown in supplementary fig. S6.

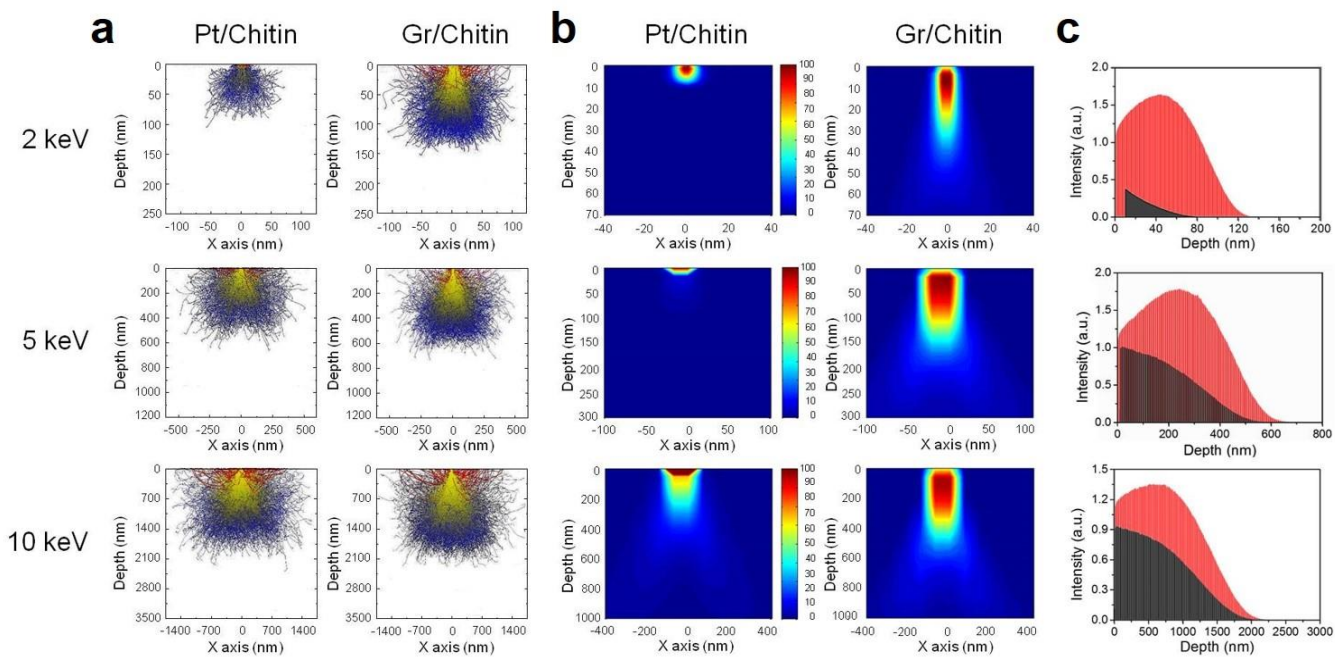

**Figure 4. Simulation results of the Pt/Chitin and the Gr/Chitin with 2, 5, and 10 keV accelerating voltages by Monte Carlo Calculation. a**, Electron trajectories of Pt/Chitin and Gr/Chitin. Blue lines are the trajectories of electron absorbed in the samples. Red lines are the trajectories of back scattered electron that would escape sample surface. **b**, Energy distribution contour mapping images of Pt/Chitin and Gr/Chitin. Dark red and dark blue are 100 and 0, respectively. **c**, Phi Rho-Z X-ray absorbed intensity of carbon (Phi = X-ray generation function, Rho-Z = a way of plot generation per unit density). Red and dark grey areas show the total intensities of Gr/Chitin and Pt/Chitin, respectively. The absorbed intensity of the Gr/Chitin are 87, 290, and 470 at 2 keV, 5 keV, and 10 keV, respectively. The intensity of Pt/Chitin are 6, 116, and 259 at 2 keV, 5 keV, and 10 keV, respectively.

# Supplementary Information

## A. Materials and Methods

### 1. Preparation of monolayer graphene

Monolayer graphene was synthesized by chemical vapor deposition method on a high purity copper foil (Alfa Aesar, 99.999 %). The flows of $H_2$ and $CH_4$ were 70 and 650 mTorr, respectively. Poly(methyl methacrylate) (PMMA) was spin-coated on the as grown graphene as a support. Unnecessary back-side graphene was etched by using $O_2$ plasma. Before final use, the PMMA film was removed by hot acetone. The remaining copper foil was etched in 1.8 wt% ammonium persulfate (APS) solution, followed by rinsing with distilled water (DI-water) several times.

### 2. Electron microscopy of biological samples.

Electron microscopic imaging and chemical analysis were carried out using field-emission scanning electron microscope (FE-SEM, Carl Zeiss, SUPRA 55VP) equipped with an EDS system that operates at 2 to 20 keV. The resolution limit is 1.0 nm at 15 keV and 1.7 nm at 1 kV, respectively. The pressure in the chamber during observation was less than $10^{-5}$ mbar. All the experiments were carried out at room temperature. The sputter coater (BAL-TEC, SCD 005) was used for preparation of Pt-coated samples under argon atmosphere. The applied current and working pressure of the chamber were around 20 mA and $5 \times 10^{-2}$ mbar, respectively. The sputtering time was 150 seconds, and the work distance was 50 mm for 10 nm-thick Pt-coating.

### 3. Preparation of water fleas fed on cerium oxide nanoparticles

The water fleas, *Daphnia pulex*, were purchased from Green Fish Mall. The ~25 nm $CeO_2$ nanoparticles were purchased from Sigma-Aldrich and dispersed in a natural water with 5 mg/100 ml concentration. One day after putting in 50 ml $CeO_2$ solution, the water fleas were rinsed for a few minutes to remove residual salts and $CeO_2$ nanoparticles, followed by natural drying.

### 4. Monte Carlo simulation by CASINO software

To confirm the possibility of graphene sheet as a membrane for electron microscopy of non-conducting biological objects, CASINO v.2.48, a modelling software based on Monte Carlo simulation, was used in this work (http://www.gel.usherbrooke.ca/casino/). The graphene layer was configured as an 1 nm thick carbon layer. Chitin was configured as a 10,000 nm thick layer consisted of C, O, N, and H with ratios of 0.4, 0.25, 0.05, and 0.3, respectively. The number of electrons to simulation and displayed trajectories were 100,000 and 2,000, respectively. As a physics model, the Mott by interpolation was selected in a total cross section and a partial cross section. The Casnati model was selected in effective section ionisation, and the model by Joy and Luo (1989) was selected in an ionisation potential. The Press model was chosen as a random number generator, and the Drouin's model (1996) was selected as a directing parameter.

## B. Supplementary Figures

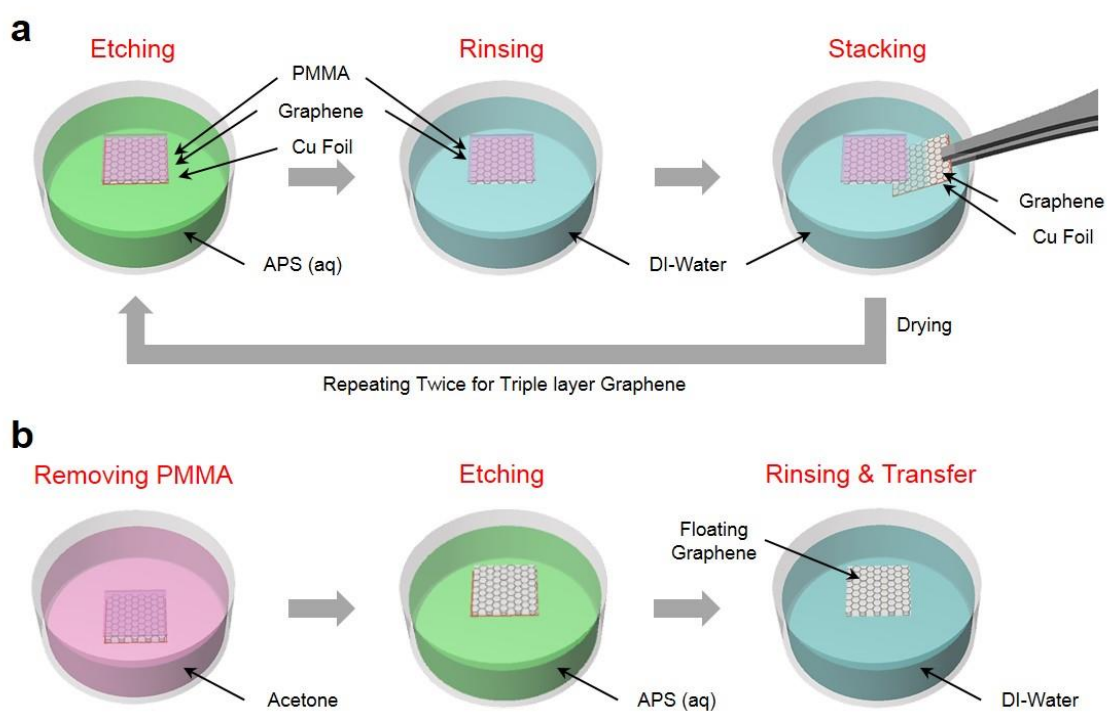

**Figure S1. a** and **b,** Schematic diagram showing the processes to prepare multilayer graphene films, which includes Cu etching, rinsing, and multi-stacking. After removing PMMA and etching Cu, the multiply stacked graphene layer is ready to be used for biological sample coating for EM analysis.

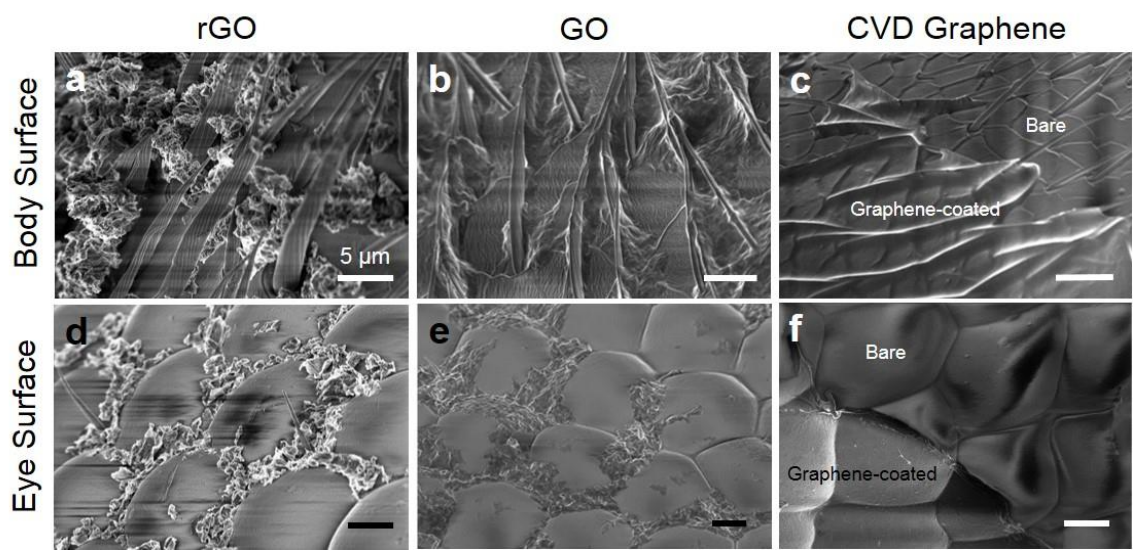

**Figure S2.** Comparison between reduced graphene oxide (rGO), graphene oxide (GO), and CVD graphene coatings for SEM imaging. **a-c**, FE-SEM images on the hairy body surface of an ant coated with rGO, GO, and 3-layer CVD graphene, respectively. Scale bars, 5μm. The rGO and GO coated samples were prepared by dipping in 0.1 wt% rGO and GO solution for 24 hours, respectively. Both rGO and GO flakes didn't cover on the sample surface, resulting in heavy charging on the surface during SEM observation. **d-f**, FE-SEM images of ant's eyes coated with rGO, GO, and 3-layer CVD graphene, respectively. Scale bars, 5μm. Acceleration voltages, 2 keV.

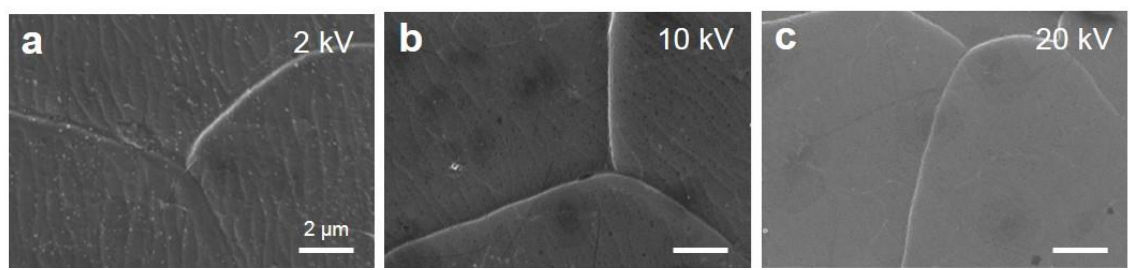

**Figure S3.** FE-SEM images of the graphene-coated ant (gaster) with increasing acceleration voltage from 2 to 20 keV. No damage was observed even at 20 keV. Scale bars, 2 μm.

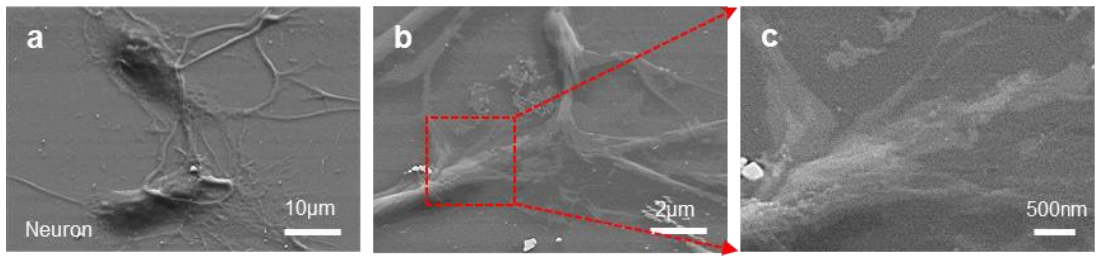

**Figure S4. a-c,** FE-SEM images of the graphene-coated hippocampal neuron with only glutaraldehyde fixation at 2 keV. Neuron cells were cultured on the graphene-coated Si wafer. The dendrites and growth cones of hippocampal neuron cells with graphene coating are visualized by SEM, which are only fixed by glutaraldehyde without other processes of animal cells for SEM analysis such as dehydration and staining. We expect that further combination with microfluidic methods would enable the real-time live imaging of bacteria and cells by electron microscopy in the future. Scale bars, 10 μm, 2 μm, and 500 nm, respectively.

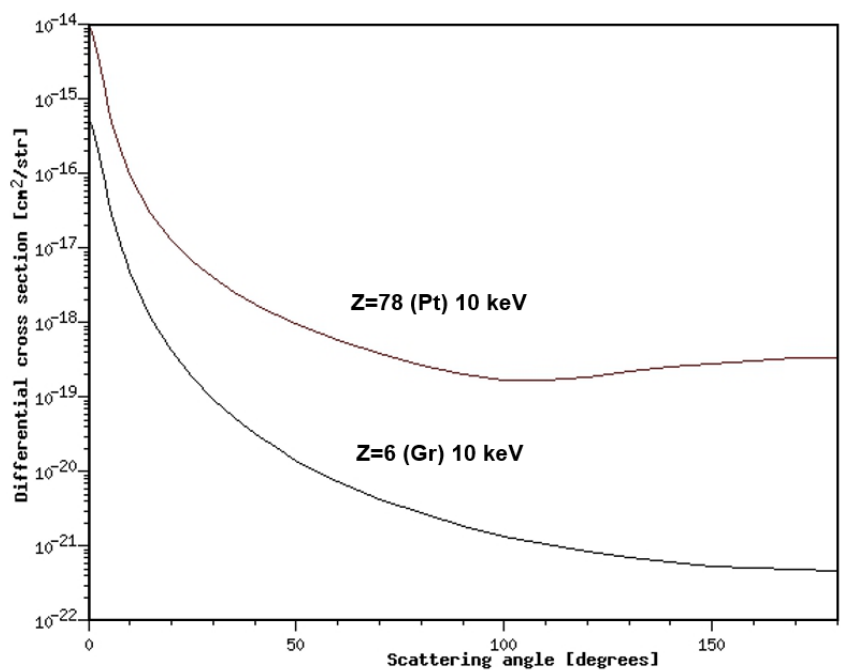

**Figure S5.** Differential cross section of incident electrons (10 keV) with respect to scattering angles for carbon (Z=6) and platinum (Z=78). Ref. Electron Scattering in Solids (http://www.ioffe.rssi.ru/ES/)

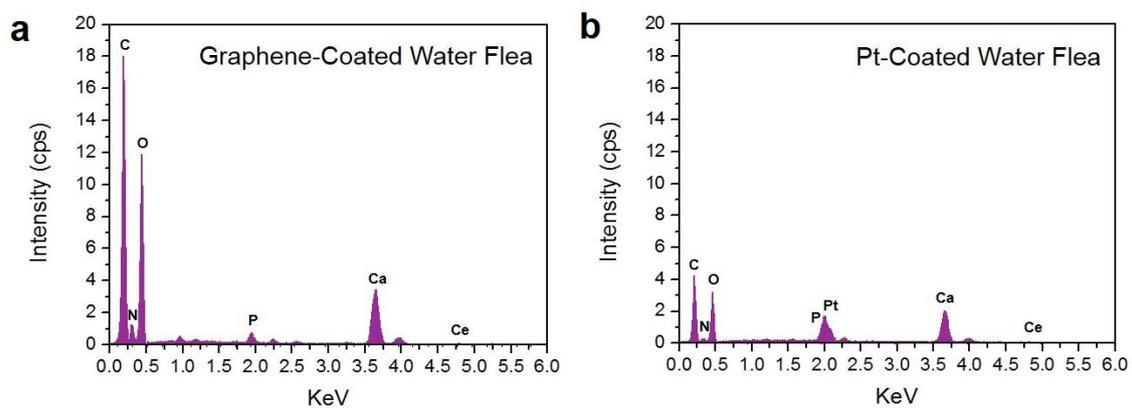

**Figure S6. a** and **b,** EDS spectra of graphene-coated and Pt-coated water fleas fed on $CeO_2$ nanoparticles, respectively. Acceleration voltages, 20 keV.

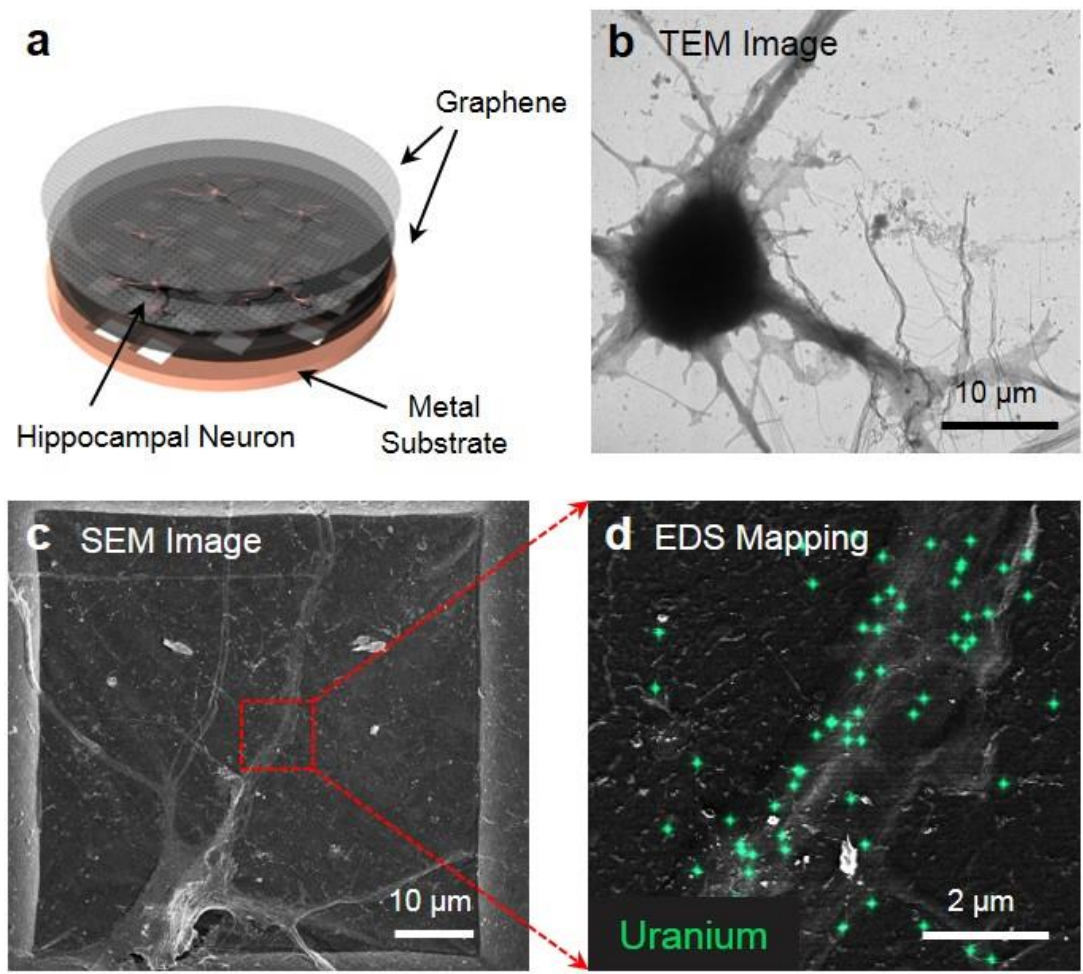

**Figure S7. SEM, TEM, and EDS mapping analysis of graphene-coated hippocampal neuron with uranyl acetate staining.** Neuron cells were cultured on the TEM grid coated with graphene, followed by fixation, dehydration, and uranyl acetate staining for positive stain in electron microscopy. After that, another graphene was transferred on the top of the neuron sample, followed by drying completely and keeping in a desiccator. **a**, Schematic image of the hippocampal neuron with 3-layer graphene coating. **b** and **c**, TEM and SEM images of the graphene-coated neuron cell. The dendrites of the neuron were clearly displayed by electron microscopy. Scale bars, 10 µm. **d**, Representative SEM and EDS mapping images of graphene-coated dendrite with uranly acetate staining of (**c**). The uranium bound to cell membrane is vividly visualized by uranium-selective EDS mapping of the graphene-coated neuron sample. Accelerating voltage 20 keV, Scale bar, 2 µm.

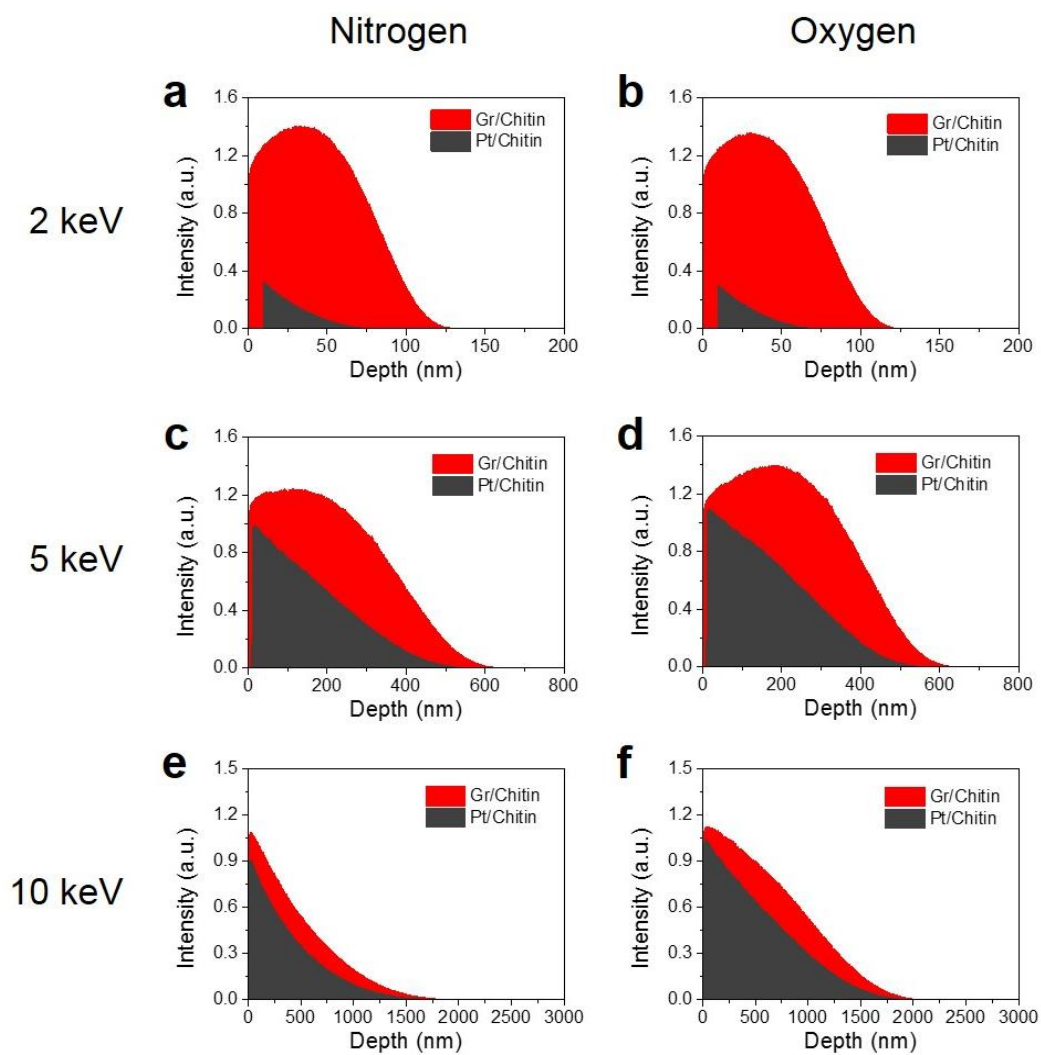

**Figure S8.** X-ray intensities of nitrogen (**a**, **c**, and **e**) and oxygen (**b**, **d**, and **f**) calculated by Monte Carlo simulation with different accelerating voltages. The red and grey areas show the absorbed X-ray intensity by Gr/Chitin and Pt/Chitin, respectively. The intensities of nitrogen in Gr/Chitin and Pt/Chitin are 11/31/29 and 0.74/14/19 at 2/5/10 keV, respectively. The intensities of oxygen in Gr/Chitin and Pt/Chitin are 83/338/452 and 5/162/312 at 2/5/10 keV, respectively.

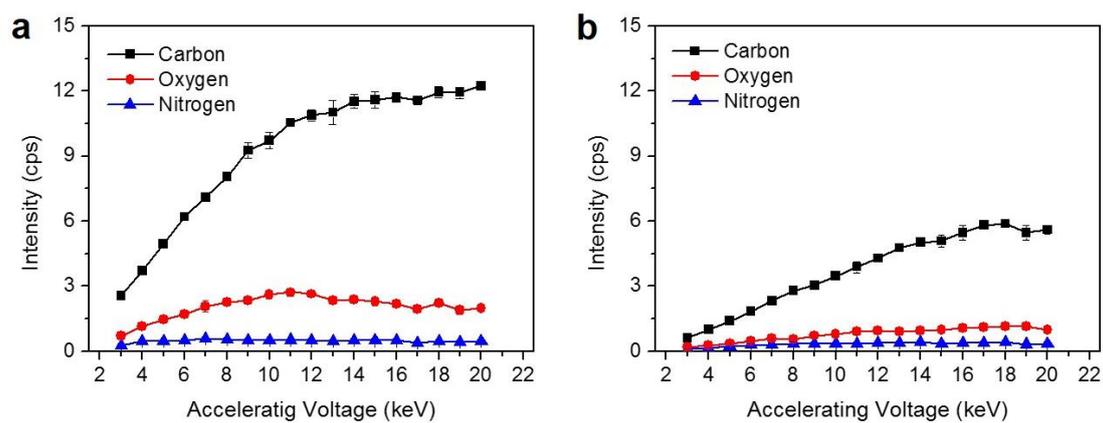

**Figure S9.** EDS signal intensities of graphene-coated (**a**) and Pt-coated ant (**b**) with increasing acceleration voltages from 3 keV to 20 keV. Working distance and accumulation time were 8.5 mm and 40 seconds, respectively.